# Transmission x-ray microscopy at low temperatures – analyzing supercurrents with high spatial resolution

*J. Simmendinger[1], S. Ruoss[1], C. Stahl[1], M. Weigand[1], J. Gräfe[1], G. Schütz[1] and J. Albrecht[2,*]*

[1] Max Planck Institute for Intelligent Systems, Heisenbergstr. 1, D-70569 Stuttgart, Germany
[2] Research Institute for Innovative Surfaces FINO, Beethovenstr. 1, D-73430 Aalen, Germany

Scanning transmission x-ray microscopy has been used to image electric currents in superconducting films at temperatures down to 20 K. The magnetic stray field of supercurrents in a thin YBaCuO film is mapped into a soft-magnetic coating of permalloy. The so created local magnetization of the ferromagnetic film can be detected by dichroic absorption of polarized x-rays. To enable high-quality measurements in transmission geometry the whole heterostructure of ferromagnet, superconductor and single-crystalline substrate has been thinned to an overall thickness of less than 1 µm. With this novel technique local supercurrents can be analyzed in a wide range of temperatures and magnetic fields. A magnetic resolution of less than 100nm together with simultaneously obtained nanostructural data allow the correlation of local supercurrents with the micro- and nanostructure of the superconducting film.

Scanning transmission x-ray microscopy (STXM) is a highly desirable tool for the investigation of materials at small length scales[1]. In particular, microscopy with polarized x-rays provides the access to magnetic ordering phenomena via dichroic absorption (x-ray magnetic circular dichroism, XMCD) and gives rise to outstanding possibilities[2]. To demonstrate the power of this approach we analyze the electric current distribution in thin superconducting films at low temperatures. A full understanding of electric transport mechanisms in high-temperature superconductors is a prerequisite towards reliable and efficient applications of such complex materials. Promising information can be gathered by imaging magnetic structures with highest spatial resolution[3]; in best cases resolving individual vortices in real, application-ready materials. Imaging superconducting vortices under laboratory conditions can be achieved with several methods, commonly with scanning SQUID microscopy[4], magnetic force microscopy[5] and scanning Hall microscopy[6]. Single vortices have also been detected by microscopy based on nitrogen-vacancies[7] or scanning tunneling spectroscopy[8]. A very successful technique to visualize magnetic structures in superconductors is magneto-optical imaging with visible light via a ferromagnetic sensor layer[9,10]. Here, flux pinning and flux dynamics can be studied over a wide range of the phase diagram with respect to temperature and magnetic field. However, the spatial resolution is limited and the structural information is not obtained. In this work we present a novel method based on polarized x-rays exhibiting analogue advantages to the magneto-optical technique but provides information on the nanostructure as well. In addition, the spatial resolution can be dramatically improved.

Two severe difficulties arise from the large absorption coefficient of soft x-rays in materials and the complicated demands on the experimental set-up to access temperatures different from ambient conditions, in particular low temperatures. Both issues have to be overcome when addressing magnetic phenomena in transmission geometry that occur in these films at low temperatures. Using the magnetic field distribution of a current carrying superconducting film as example we are able to show that low temperature magnetic transmission x-ray microcopy is able to access the magnetic signal generated by the superconductor. In particular, we can measure the stray field distribution of the superconductor in broad areas of the magnetic field vs. temperature phase diagram with a spatial resolution of 85nm.

To exploit the XMCD effect for microscopic imaging a ferromagnetic sensor layer is utilized to detect the local magnetic field arising from supercurrents. Since the magnetic interaction between superconducting and ferromagnetic materials has many aspects[11] we focus here on systems where the interaction is dominated by magnetic dipolar coupling[12–14]. A highly adapted system of matching superconducting and magnetic properties provides the basics for imaging the flux density pattern via soft x-rays[15]. Thus, we use the well-studied high-temperature superconductor $YBa_2Cu_3O_{7-\delta}$ (YBCO) in conjunction with a soft ferromagnetic material such as permalloy ($Ni_{0.8}Fe_{0.2}$, Py). Py is our material of choice due to its small in plane coercive field of about 13 Oe which is constant within a range of 1 Oe in the temperature range from 5 to 100 K (see Fig. 1).

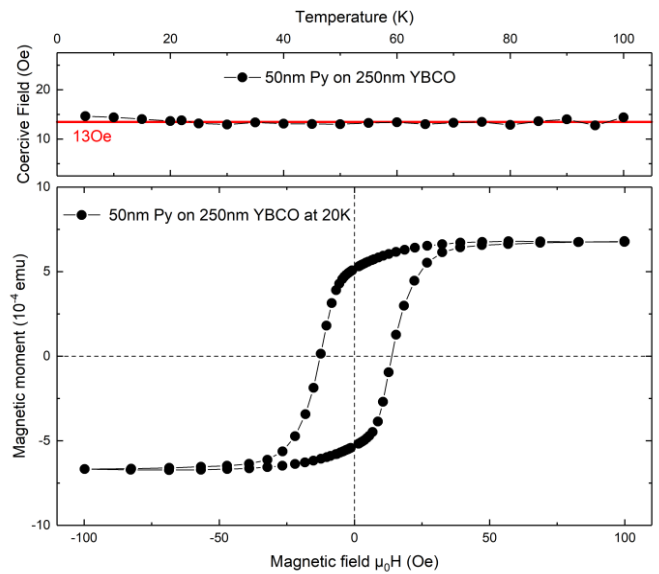

*Figure 1:* (Bottom) SQUID hysteresis loop in the plane of 50 nm Py on 250 nm YBCO measured at T = 20 K. (Top) The

*variation of the coercive field $\mu_0 H_c$ (red line) is less than 1 Oe in the temperature range from 5 to 100 K.*

The typical magnetic flux density to map is more than one order of magnitude larger as the Py coercivity and is therefore mapped into the ferromagnetic sensor layer. Its magnetization distribution can then be measured by scanning x-ray microscopy (SXM) in the total electron yield mode[16,17] or by photoemission electron microscopy (PEEM)[18].

In this work we report on results that are obtained in a transmission set-up leading to an excellent signal to noise ratio. In contrast to surface sensitive approaches, such as TEY or PEEM, STXM captures the superconducting responds of the whole sample volume. Thus, concerns of overestimating structural defects or even not depicting the actual superconducting state[19,20] are counteracted. Furthermore, relaxation effects of the flux density distribution can be switched off due to the possibility to access data in finite magnetic fields. On the other hand time-dependent relaxation effects can be addressed in detail with our STXM approach.

We use thin, optimally doped YBCO films epitaxially grown on single crystalline SrTiO$_3$ (001) (STO) with Py as a sensor layer on top. To reduce the pair-breaking effect of the ferromagnet a 5 nm STO protection layer has been put in between which limits the interaction to dipolar magnetic coupling[21]. Finally, 2 nm of aluminum are deposited on top.

The x-ray microscopy measurements are performed at the STXM set-up MAXYMUS at Bessy II in Berlin. It has been upgraded with a liquid Helium cryostat to be the first STXM reaching temperatures down to 20K. The microscope is able to record XMCD images with high spatial resolution of less than 30 nm[1,22], on short time scales of nanoseconds and below[23,24] as well as element specific and biological contrast[25]. In case of the transmission set-up used in this work the access of a thin perovskite layer is very demanding since the growth on a matching single crystalline substrate is necessary. The separation of the materials from the substrate is challenging but has been reported possible via an etching method[26]. Our approach is to remove the substrate after the growth process by focused ion beam (FIB) milling[27] to a thickness of about 1 µm and below. Only with the substrate thinned to this extent transmission of soft x-rays becomes possible. In the following we report on the first successful transmission measurement of superconducting transport properties in the soft x-ray regime.

The elaborate sample geometry that allows the realization of a transmission experiment at epitaxial oxide layers is depicted in Figure 2.1. For simplicity the protection layer (STO) and the final aluminum capping are not displayed.

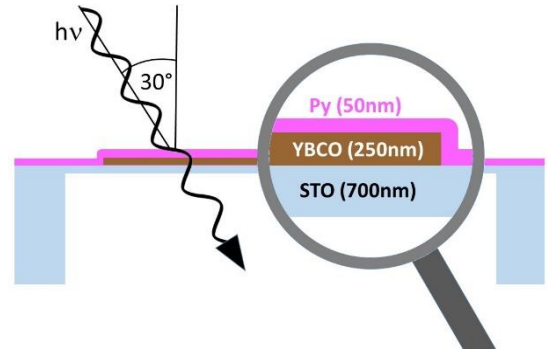

***Figure 2.1:*** *Cross-sectional sketch of the heterostructure used in this work: An epitaxial YBCO film (brown) is grown on single-crystalline STO (001). After patterning the YBCO film to 20 x 20 µm$^2$ square a Py film (pink) is grown on top. To enable transmission of x-rays under an incident angle of 30° the substrate underneath is thinned by focused ion beam irradiation until a complete thickness of the order of 1 µm is reached.*

This kind of transmission samples can be produced following a series of educated preparation steps that are listed in the following. First, an epitaxial film of optimally doped YBCO is grown by pulsed-laser deposition (PLD) on a single-crystalline SrTiO$_3$ (001) substrate with a thickness of $d_{STO}$ = 500 µm. The thickness of the YBCO film is 250 nm, a superconducting transition at $T_c$ = 88 K is found when measuring the diamagnetic signal. The film exhibits a high critical current density of $j_c$ > 3 x 10$^{11}$ A/m$^2$ at T = 10 K. These so prepared thin films are then structured to 20 µm squares using photo lithography and ion beam etching with Argon ions. Therefore, the samples were coated with a UV-activated coating containing inorganic Silica groups to resist the ion-beam etching process. The coating has been exposed using the "Kloé Dilase 250" Laserwriter with a Laser wavelength of 375 nm. After pulsed ion beam etching and removal of the remaining coating, 50 nm of Py and 2 nm of aluminum are deposited on the sample using RF sputtering at room temperature. To avoid diffusion of Fe and Ni, a thin STO interlayer (d = 5 nm) between Py and YBCO is introduced. No change of the superconducting properties ($T_c$ and $j_c$) before and after Py deposition is found.

Subsequently, the samples are cut into pieces of 1 x 1 mm$^2$ and the 500 µm thick STO substrate is thinned to approximately 20 µm using conventional mechanical thinning to achieve a plan-parallel surface. Then, the STO substrate underneath the YBCO structures is further thinned to a thickness $d_{STO}$ < 1µm using a "FEI Nova 600 NanoLab" focused ion beam (FIB) source using Gallium ions at 30 kV with a beam current of 20 nA.

Figure 2.2 shows a scanning electron micrograph of the samples used for the transmission experiments. The thinned area is seen rectangular region in the center each panel, respectively.

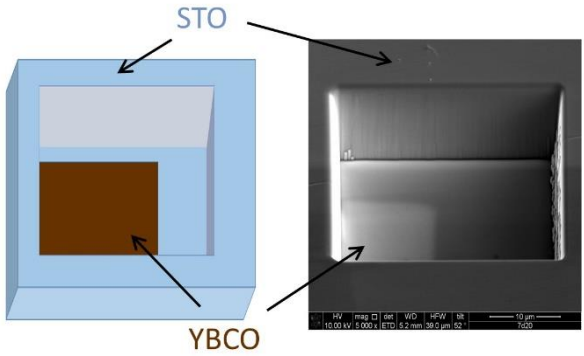

*Figure 2.2:* SEM micrograph of the backside-thinned strontium titanate (STO) substrate. Inside the thinned rectangular area the superconducting YBCO film can be seen through the remaining STO. The sketch on the left shows the substrate with the thinned area, the YBCO film is on the back.

Inside the thinned window a rectangular bright area appears in the electron micrograph. This indicates the position of the superconducting film on the back that can be seen through the remaining thin STO layer.

First, the structural properties of the superconducting film are characterized. The surface of the superconductor-ferromagnet bilayer is investigated both by scanning electron microscopy (SEM) and x-ray absorption microscopy (SXM). The corresponding results are seen in Figure 3.

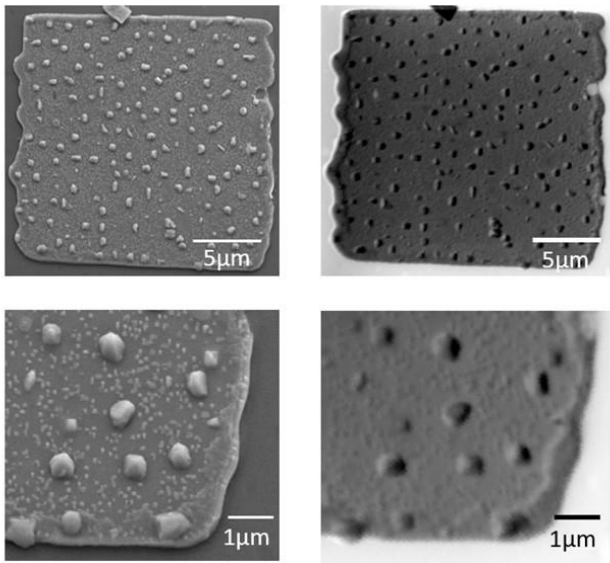

*Figure 3:* Surface of the superconducting film characterized by scanning electron microscopy (left column) and x-ray absorption microscopy (right column).

Both experimental results reveal analogue information. We find a complex microstructure consisting of large particles with a typical diameter of about 0.5 µm and much smaller (rectangular-shaped) structures on a length scale of about 100 nm visible as small white spots in the image at the bottom left. The large features refer to a-axis oriented outgrowths in the superconducting film[24], exceeding the average film thickness of 250 nm by several 100 nm. The small structures correspond to the ubiquitous growth islands in epitaxially grown YBCO films. The found surface topography is conserved during deposition of 50 nm Py and 2 nm Al.

Figure 4 depicts the result which are obtained with the presented transmission x-ray microscopy at a 20 µm x 20 µm YBCO square at a temperature of T = 20 K. The superconductor has been prepared into the remanent state by applying a magnetic field of $\mu_0 H$ = 200 mT and subsequent reduction to zero. In this case the supercurrents are trapped in the superconductor and form four domains of constant current density. This electric current distribution lead to in-plane components of the magnetic flux density at the surface of the superconductor that are oriented towards the edge of the film. The magnetic (XMCD) signal is defined as the difference between left and right circular polarized x-rays at the Ni $L_3$ edge in % of the XAS signal[28].

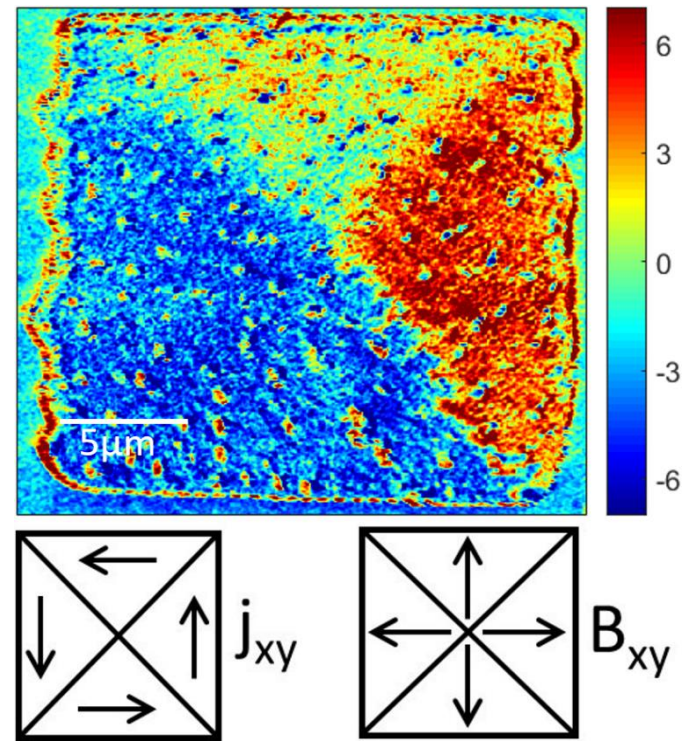

*Figure 4.1:* Image of a superconducting YBCO thin film in the remanent state obtained by magnetic x-ray microscopy at T = 20 K. The color scale represents the local strength of the XMCD signal in %. There is a roof-shaped pattern consisting of four congruent triangles. This is related to local flux and current distributions illustrated in the corresponding sketches below.

The colorscale map of the superconducting film obtained by low-temperature transmission x-ray microscopy clearly allows the identification of the four triangular current domains via their magnetization of the soft-magnetic sensor layer. It is possible to separate individual current domains via their stray fields. The magnetic resolution is basically given by the domain wall width of the used ferromagnet[17], namely 85nm for a 90° and 142nm a for 180° domain wall in good agreement with literature values for Py[29,30]. The irregular, defect-like structures that are dispersed inside the triangular domains refer to a structural contrast that is originated by the growth islands and misoriented outgrowths in the YBCO film. These have been already addressed in Figure 3[28].

The correlation between microstructure and magnetic properties is displayed in more detail in Figure 4.2.

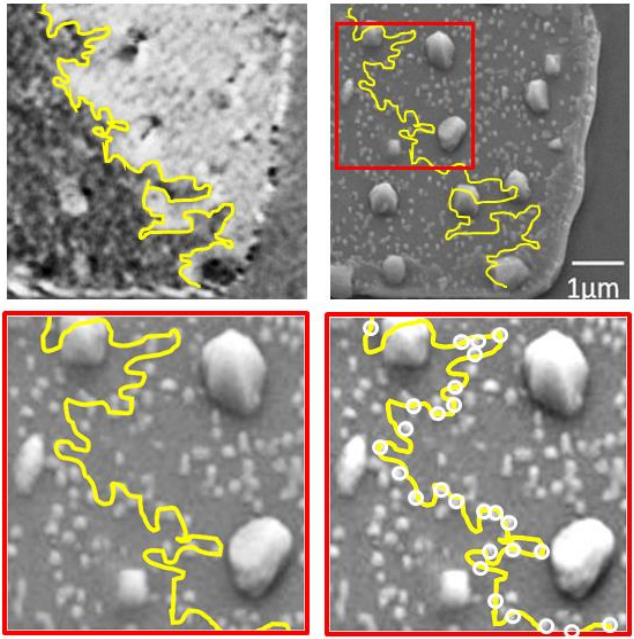

*Figure 4.2: The magnetic domain wall separating the bottom and the right current domain of Figure 4.1 is marked with a solid yellow line (top left). The yellow line is transferred into the electron micrograph (top right). The bottom panels show the rectangular section (red solid line). The image on the bottom left clearly shows the close correlation of the domain wall position and the small growth islands. For clarity the position of islands in vicinity of the domain wall are highlighted (bottom right).*

The first panel of Figure 4.2 shows the position of the magnetic domain wall that separates two areas of different magnetic contrast, e.g. two current domains, as yellow line. The image displays a section of Figure 4.1 extracted at the edge on the bottom right. This yellow line is transferred into the equivalent section of the scanning electron micrograph (top right). Here, it is seen that the position of the domain wall is not directly influenced by the large outgrowths in this section the film. The domain wall seems to follow the locations of the smaller structures being growth islands of YBCO. This becomes more obvious in higher magnification (bottom images). The domain wall is spanned directly between the bright structures with sizes of about 100nm. For clarity, the structures that match the position of the domain wall are marked by small white circles (bottom right). This panel allows the identification of the dominant kind of structural defect which is responsible for the pinning properties of this YBCO thin film. It can clearly be stated that the large, 500nm-sized outgrowths do not contribute substantially.

The magnetic signal-to-noise ratio of this novel technique is large enough to provide a significant dichroic signal for systematical investigations on the behavior of the superconducting thin film with respect to changing temperatures and magnetic fields.

As an example of these enormous possiblities Figure 5 shows the magnetic flux density distribution inside the superconducting YBCO thin film in varying external magnetic fields obtained by transmission x-ray microscopy at T = 20K.

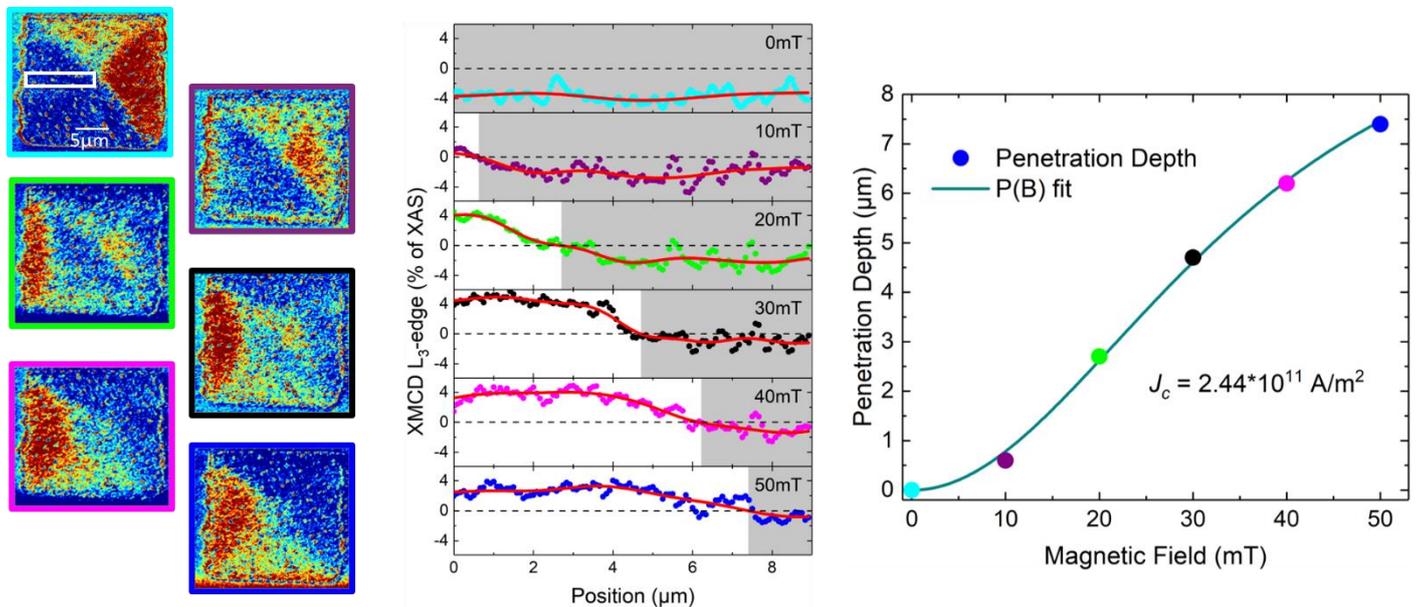

*Figure 5: (Left) Change of the magnetic flux density pattern inside the YBCO thin film caused by variation of the external magnetic field at T=20K. (Middle) The profiles are extracted at the white rectangular area; red solid lines refer to an averaging of the data. Starting in the remanent state (0mT) an increasing external field (up to 50mT) reverses the electric screening currents in the superconductor (white background) and the border region (penetration depth) is further shifted towards the center of the superconductor.(Right) Symbols represent the penetration depth as a function of external magnetic field with a fit according to Eq.(1). Errors are within the size of the symbols.*

The panels in the left part of Figure 5 show the development of the dichroic pattern when changing the external magnetic field visualized by magnetic transmission x-ray microscopy.

The image sequence starts in the remanent state at $\mu_0 H_{ex}$ = 0 mT (top image) and runs in 10 mT steps up to 50 mT. A systematic rearrangement of the domain pattern is observed, red and blue areas change their sign when areas of opposite contrast enter from the edge of the film. This refers to a magnetization reversal caused by the inversion of the electric screening currents in these areas originated by the opposite field sweep (down sweep to remanence and up again as visualized) and the corresponding induction phenomenon.

Quantitative information is provided in the mid-panel. It displays profiles extracted at the rectangular area on the left part of the superconducting film at each external field. The transition region between areas of different orientation is moving monotonously with sweeping the external field. This is particularly marked inside the displayed profiles as the background color switches from red to blue at the position of the sign change. In this representation the propagation of the transition line can be analyzed with highest spatial accuracy. In addition, we find that the line of the magnetization reversal is rough. This is attributed to the inhomogeneities in the film that have been discussed before. The non-existing correlation between the distribution of the large misoriented grains and the domain pattern well coincides with previous findings[28].

We now turn to a quantitative investigation of the superconducting properties, namely the superconducting current density which we extract by fitting the penetration depth $P(B_a)$ as function of the external magnetic field utilizing Eq.(1)[31]

$$P(B_a)[\mu m] = (A - A/[cosh(B_a/B_c)])/2 \quad (1)$$

$$B_c \propto J_{c,PenDepth} * d_{YBCO} * \mu_0$$

where $d_{YBCO}$ is the thickness of the YBCO film, $A$ the size of the square and $\mu_0$ is the vacuum permeability.

The right panel of Figure 5 depicts the penetration depth extracted from the position of the sign change in the XMCD profiles versus the external magnetic field. The color of the data point corresponds to the color of the profile and the solid line is a numerical fit of the data using Eq. (1). The numerical fit describes the experimental data with high accuracy; the critical current density can be determined to be $J_{c,PenDepth}$=2.44 A/m² at T=20K.

To correlate the latter approach with complementary results we plot the temperature evolution of $J_{c,PenDepth}$ for the 20 x 20 µm² squares (cyan stars) in Figure 6 together with $J_{c,SQUID}$ of the unstructured YBCO film (cyan dashed line), obtained by non-local SQUID magnetometry, and the local XMCD values (diamonds).

Except for a small offset between SQUID and XMCD data the two approaches are in good agreement. A typical temperature dependence of the supercurrent density of a thin YBCO film is found[32]. The offset can be explained by pair-breaking effects in the top few nanometers of the superconductor caused by the inverse proximity effect and the sputtering process of the Py sensor layer[15,16].

Note $J_{c,SQUID}$ and $J_{c,PenDepth}$ are non-local approaches and can only be evaluated for some defined shapes employing a model for current densities in high-$T_c$ superconductors. However, due to the bulk sensitive STXM approach with absolute values for the XMCD signal, our raw XMCD data can be directly converted into a local current density. It is important to stress that this a local routine with a resolution of 85nm, and furthermore, it is independent from a predefined model.

Figure 6 displays the development of the magnetic signal (cyan/red squares), created by the electric currents of the superconductor in the remanent state, in the temperature range from 20K to 100K. To obtain reliable data the magnetization of the background is subtracted to extract the pure signal of the superconductor. It has already been shown that the superconductor-induced XMCD effect decreases with increasing temperature[16]. This behavior known from absorption experiments is now found as well in the microscopy data.

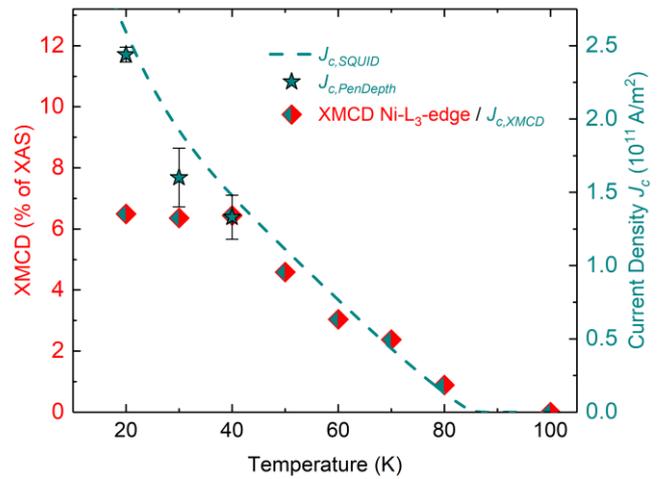

*Figure 6:* Temperature dependence of the current density $J_c$ obtained via fitting the penetration depth versus magnetic field (cyan stars) of a 20x20µm square and for the unstructured YBCO sample via SQUID magnetometrie (cyan dashed line). Both the absolute values and the temperature behavior of the two approaches coincide. On this basis the XMCD Ni $L_3$-edge signal (cyan/red squares, error within symbol size), measured in % of the XAS, can be assigned to a current density $J_c$.

We find that up to T = 40 K the XMCD signal remains constant. Here, the stray field of the superconductor saturates the ferromagnetic sensor layer. At temperatures above T = 40 K the magnetic stray field of the superconductor further decreases and falls below the threshold for saturation. Thus, the found magnetization reduces with increasing temperature.

A closer examination reveals a similar temperature dependence for $J_{c,SQUID}$ and the XMCD signal between 40 K and 80 *K* which suggests the XMCD signal to be a direct measure for the local current density ($J_{c,XMCD}$) in this temperature range. This finding is supported by the matching values of $J_{c,XMCD}$ and $J_{c,PenDepth}$ at 40K. Hence, the XMCD signal is proportional to the current density $J_c$, since it can be calibrated via both $J_{c,SQUID}$ and $J_{c,PenDepth}$.

The finite XMCD signal at 80 K in combination with the coercive field (13 Oe) of the Py sensor layer enables an estimation for the sensitivity of our STXM approach. MOFE measurements performed at an unstructured YBCO sample, revealed the in-plane component of the magnetic stray field to be 20 Oe at 80 K. Thus, the threshold for the STXM sensitivity lies in the range from 13 Oe to 20 Oe.

Finally, above T = 90 K the superconductor is in the normal state and the ferromagnet relaxes to remanence.

In conclusion, we have shown that transmission magnetic x-ray microscopy is able to analyze the electric current transport in epitaxial superconducting films. The combination of the first liquid helium cryostat in a scanning x-ray microscope with an educated fabrication of ultra-thin single-crystalline heterostructures enables the measurement of the local stray field distribution in the superconducting state. Our method combines parallel acquisition of magnetic and structural information, highest spatial resolution and brilliant signal-to-noise ratio. In addition, the magnetic signals are accessed quantitatively without the need of additional image calibration routines. This provides access to a quantitative analysis of the superconducting current density with a resolution of 85nm in the phase diagram with respect to temperature and magnetic field. In combination with structural information with a spatial resolution on 30 nm our method represents a unique tool for the analysis of magnetic low-temperature phenomena in general.

The authors are grateful to G. Christiani, U. Eigenthaler, B. Stuhlhofer and M. Kelsch for their contribution to the fabricating of the samples and P. van Aken for his support.